\documentclass[12pt,epsfig]{article}

\usepackage[dvipsnames,usenames]{color}
\usepackage{graphicx}
\usepackage{amssymb}
\usepackage{amsmath}
\usepackage{epsfig}
\usepackage{amssymb}
\usepackage{graphicx}
\usepackage{graphics}

\def\mathnew{\mathsurround=0pt}
\def\simov#1#2{\lower .5pt\vbox{\baselineskip0pt \lineskip-.5pt
       \ialign{$\mathnew#1\hfil##\hfil$\crcr#2\crcr\sim\crcr}}}
\def\beq{\begin{equation}}
\def\enq{\end{equation}}

\def\Mesz{M\'esz\'aros~}
\def\bitm{\bibitem}

\def\msun{M_\odot}

\def\bec{\begin{center}}
\def\enc{\end{center}}

\def\swift{\textit{Swift~}}
\def\fermi{\textit{Fermi~}}

\def\gbm{\textit{GBM~}}
\def\lat{\textit{LAT~}}
\def\vareps{\varepsilon}

\def\cm{\hbox{~cm}}

\def\GeV{\hbox{~GeV}}

\begin{document}
\Large
\centerline{\bf Gamma Ray Bursts}
\normalsize
\centerline{\bf Peter M\'esz\'aros}
\centerline{Center for Particle and Gravitational Astrophysics}
\centerline{Dept. of Astronomy \& Astrophysics and Dept. of Physics}
\centerline{Pennsylvania State University, University Park, PA 16802, USA}

\begin{abstract}
Gamma-ray bursts have been detected at photon energies up to tens of GeV.
We review some recent developments in the X-ray to GeV photon phenomenology
in the light of  \swift and \fermi observations, and some of the theoretical
models developed to explain them, with a view towards implications for C.T.A.
\end{abstract}


\section{Introduction}
\label{sec:intro}

Gamma-ray bursts (GRB) are brief events occurring at an average rate of 
a few per day throughout the universe, which for a period of seconds
flood with their radiation an otherwise almost dark gamma-ray sky.
While they are on, they far outshine all other sources of gamma-rays in the
sky, including the Sun. In fact, they are the most concentrated and brightest
electromagnetic explosions in the Universe. Since the discovery in 1997 by the
{\it Beppo-SAX} satellite of GRB X-ray afterglows, which enabled ground-based 
telescopes to detect their optical counterparts, we know that these objects 
are at cosmological distances. Thanks mainly to the subsequent \swift satellite, 
we now have detailed
multi-wavelength data for many hundreds of bursts, and redshifts for over 150.
In fact, the GRB prompt electromagnetic energy output during tens of seconds 
is comparable to that of the Sun over $\sim {\rm few} \times 10^{10}$ years, 
or to that of our entire Milky Way over a few years; and their X-ray afterglow
over the first day after the outburst can outshine the brightest X-ray quasars. 

The current interpretation of this spectacular phenomenon is that about a solar rest 
mass worth of gravitational energy is released in a very short time (seconds or less) 
in a small region of the order of tens of kilometers by a cataclysmic stellar event.
The latter, in some cases, is the collapse of the core of a massive star, or in some 
cases may be the merger of two compact stellar remnants, ultimately leading to a black 
hole.  The mainstream GRB scenario envisions that only a small fraction of this energy 
is converted into electromagnetic radiation, through the dissipation of the kinetic 
energy of a collimated relativistic outflow, a ``fireball" with bulk Lorentz factors  
of $ \Gamma \gtrsim 300$, expanding out from the central engine powered by the 
gravitational accretion of surrounding matter into the collapsed core or black hole. 

The most often invoked mechanisms for generating the observed non-thermal photons 
are synchrotron radiation and/or inverse Compton (IC) scattering by relativistic
electrons which have been accelerated to a power-law distribution in the shocks 
expected in the optically thin regions of  the outflow. These may be internal shocks, 
resulting in prompt $\gamma$-ray emission, and also external shocks at the termination 
of the relativistic outflow, which can explain many of the properties of the afterglows.  
Other mechanisms considered for the prompt emission are, e.g., magnetic dissipation or 
reconnection in the outflow, jitter radiation in shocks, or dissipative effects
in the photosphere where the outflow transitions to optical thinness. 

More recently the \lat instrument onboard the \fermi spacecraft,  extending previous 
preliminary observations by {\it EGRET} on the Compton {\it CGRO} satellite, has shown 
that a substantial fraction of GRBs have photon spectra which extend at least to tens 
of GeV \cite{Gonzalez+03,Band+09,Kaneko+08}. 
These could be due either to leptonic mechanisms such as mentioned above,
or perhaps could be related to hadronic effects. The uncertainty about the mechanism
is, in part, due to lack of knowledge about two important parameters of the outflow. 
These are the baryon load of the outflow, and the magnetic ratio $\sigma$ between 
magnetic stresses and kinetic energy, which affect not only the bulk dynamics but 
also the possibility of accelerating protons in the shocks or the dissipation region. 
Accelerated protons could lead in principle to GRBs being as (or perhaps more) luminous 
in cosmic rays and neutrinos than in the commonly observed sub-MeV electromagnetic 
channels, in addition to predicting  significant secondary GeV and TeV photon fluxes. 

\section{GRB Phenomenology: MeV to Multi-GeV}
\label{sec:gevobs}

Soon after its launch in late 2008, \fermi started detecting GRBs with both its 
instruments, the Large Area Telescope (\lat, 20~MeV to $>300$~GeV) and the 
Gamma-ray Burst Monitor (\gbm, 8~keV to 40~MeV), which together cover  more than 
seven decades in energy. While in its first 2.5 years the \gbm has triggered on 
bursts at a rate of about 250 yr$^{-1}$, of which on average $\sim 205$ are long 
bursts and $\sim 45$ are short bursts, the \lat detected $\sim 25$ bursts in the 
same period at $\ge 100$ MeV, i.e. $\sim 9$ yr$^{-1}$; or, in the same period at
$\ge 1$ GeV it detected $\sim 12$, i.e. $\sim 5$ yr$^{-1}$ 

The brightest \lat bursts as of September 2011, 
GRB~080916C~\cite{Abdo+09_GRB080916C},
GRB~090510~\cite{Abdo+09_GRB090510,Ackermann+10_GRB090510},
GRB~090902B~\cite{Abdo+09_GRB090902B}, and
GRB~090926A~\cite{Ackermann+11_GRB090926A},
have yielded hundreds of $\gtrsim 100$ MeV photons each, and together with
the lower energy \gbm observations have yielded unprecedented broad-band
spectra. An interesting and unexpected behavior is that in many cases 
the GeV emission starts with a noticeable delay after the MeV emission. 
E.g. in  GRB 080916C, the GeV emission appears only in a second pulse, 
delayed by $\sim 4$ s relative to the first pulse (visible only in MeV);
see Fig. 1. 
Such a delay is present also in short bursts, such as GRB 090510, where it 
is a fraction of  a second. This soft-to-hard spectral evolution is clearly 
seen in all five of the bright \lat bursts above, and to various degrees 
a similar behavior is seen in other weaker \lat bursts. 
\begin{figure}[htb]
\begin{minipage}{0.8\textwidth}
\includegraphics[width=1.0\textwidth,height=3.0in,angle=0.0]{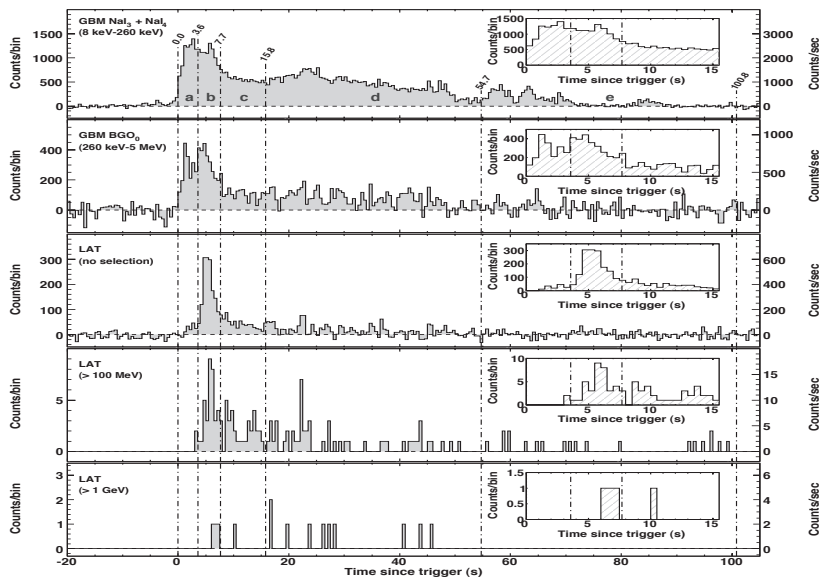}
\end{minipage}
\hspace{5mm}
\begin{minipage}[t]{0.15\textwidth}
\vspace*{-1.5in}
\caption{Light curves of GRB080916C with the \gbm (top two curves) and \lat
(bottom three curves)\cite{Abdo+09_GRB080916C}.}.
\end{minipage}
\label{fig:080916-lc}
\end{figure}

\begin{figure}[htb]
\begin{minipage}{0.75\textwidth}
\includegraphics[width=1.0\textwidth,height=3.0in,angle=0.0]{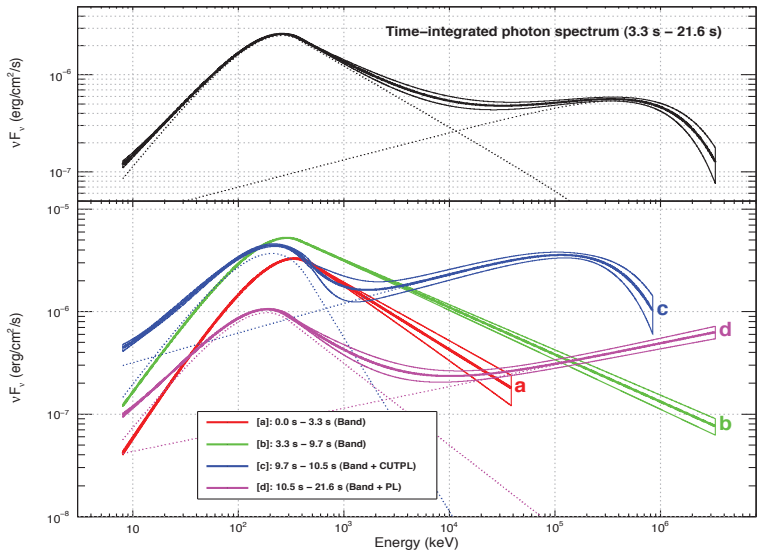}
\end{minipage}
\hspace{5mm}
\begin{minipage}[t]{0.2\textwidth}
\vspace*{-1.5in}
\caption{Spectra of GRB090926A from \fermi at four different time intervals, 
a= [0.0-3.3s], b= [3.3-9.7s], c= [9.7-10.5s], d= [10.5-21.6s] 
\cite{Ackermann+11_GRB090926A}.}
\end{minipage}
\label{fig:090926A-spec}
\end{figure}

In some burst, such as GRB 080916C and several others, the broad-band gamma-ray 
spectra consisted of simple Band-type broken power law function in {\it all} time 
bins (similar to spectrum in time bin [a] of Fig. \ref{fig:090926A-spec}).
In GRB 080916C the first pulse had a soft high energy index disappearing at GeV, while
the second and subsequent pulses had harder high energy indices reaching into the 
multi-GeV range. The absence of statistically significant evidence for a distinct 
high energy spectral component in a number of the earlier \lat bursts was puzzling, 
since such an extra component is naively 
expected from inverse Compton or hadronic effects. Instead, in each time bin a 
single broken power spectrum straddled the \gbm and (except for the earliest times) 
the \lat ranges.  The peak energy of the Band function evolved from soft to hard 
and back to soft, and in GRB 080916C as well as in other \lat bursts, the GeV 
emission persisted into the afterglow phase, typically lasting $\gtrsim 500-1000$ s. 

On the other hand, in some bursts, such as GRB090510~\cite{Abdo+09_GRB090510,
Ackermann+10_GRB090510} and GRB~090902B~\cite{Abdo+09_GRB090902B}, a second 
hard spectral component extending above 10~GeV without any obvious break 
appears in addition to common Band spectral component dominant in the lower
8 keV-10 MeV band. A similar second hard component is also present in 
GRB~090926A (Fig. \ref{fig:090926A-spec}), where in one time bin it shows a 
turnover around a few GeV. Such a second component, if present in  GRB~080916C 
and GRB~110731A, was not significant enough to claim detection.

Furthermore, in some bursts such as GRB~090510 and GRB~090902B, the same 
2nd power-law component that dominates above $\sim 100$~MeV also extends
below the Band component, and dominates below $\sim 50$~keV. This low
energy power law extension (i.e. flatter and lying below the Band component)
is also detected in at least two other bright GBM short bursts, GRB~090227B 
and GRB~090228, which have not been detected by the LAT~\cite{Guiriec+10}. 

An exciting discovery, unanticipated by {\it EGRET}, was the detection of 
high-energy emission from two short bursts (GRB~081024B~\cite{Abdo+2010_GRB081024B} 
and GRB~090510~ \cite{Ackermann+10_GRB090510}), whose general behavior 
(including a GeV delay) is qualitatively similar to that of long bursts.
The ratio of short to long GRBs is $\sim 10-20$\%, both for objects 
detected at $\ge 100$~MeV and for those detected only at $\lesssim 1$~MeV. 
However, while the statistics on short GRBs are too small to draw firm conclusions,
so far, the ratio of the \lat fluence to the \gbm fluence is $> 100\%$ for 
the short bursts as compared to $\sim 5$ -- 60\% for the long bursts. 
Thus, although fewer in number, if the acquisition is fast enough CTA might 
be able to detect short bursts.  

It is also noteworthy that, for both long and short GRBs, the $\ge 100$~MeV
emission lasts longer than the \gbm emission in the $\lesssim 1$~MeV
range. The flux of the long-lived \lat emission decays as a power law with 
time, which is more reminiscent of the smooth temporal decay of the
afterglow X-ray and optical fluxes rather than the variable
temporal structure in the prompt keV--MeV flux. This similarity in the
smooth temporal evolution of the fluxes in different wave bands has
been detected most clearly from GRB~090510~\cite{DePasquale+10,Razzaque10},
although this object also requires a separate prompt component to the \lat
emission ~\cite{He+11}.  This short burst was at $z=0.9$, and was jointly 
observed by the \fermi \lat, \gbm  and by the \swift Burst Alert Telescope (BAT), 
the X-Ray Telescope (XRT), and the UV and Optical Telescope (UVOT).  The longest-lived 
emission detected by the \fermi \lat  is $\sim 130$~minutes, from GRB~090328, longer 
than the EGRET observation of GRB~940217~\cite{Hurley+94} without any earth
occultation.

Interestingly, the \lat detected only $\lesssim 10\%$ of the bursts triggered 
by the \gbm which were in the common \gbm-\lat field of view.
This may be related to the fact that the {\it LAT}-detected GRBs, both long and 
short, are generally among the highest fluence bursts, as well as being
among the intrinsically most energetic GRBs. For instance, GRB~080916C 
was at $z=4.35$ and had an isotropic-equivalent energy of $E_{\gamma, \rm iso}
\approx 8.8\times 10^{54}$~ergs in $\gamma$ rays,  the largest ever measured
from any burst~\cite{Abdo+09_GRB080916C}. The long \lat bursts GRB~090902B
~\cite{Abdo+09_GRB090902B} at $z=1.82$ had $E_{\gamma, \rm iso}\approx 
3.6\times 10^{54}$~ergs, while GRB~090926A~\cite{Ackermann+11_GRB090926A}
at $z=2.10$ had $E_{\gamma, \rm iso}\approx 2.24\times 10^{54}$~ergs.
Even the short burst GRB~090510 at $z=0.903$ produced, within the first 2~s, an 
$E_{\gamma, \rm iso}\approx 1.1\times 10^{53}$~ergs~\cite{Ackermann+10_GRB090510}.

\section{Simple Leptonic Models}
\label{sec:gev1}

The standard internal and external shock models are the ones most commonly used in 
interpreting the \fermi data on GRBs,  as was previously the case with \swift data.
In the latter, the amount and quality of data was so large and detailed that
various extensions and refinements to the simpler standard model were considered,
and the need for this is becoming apparent also for \fermi analyses, even though, 
particularly for the \lat, the statistics are only gradually building up.

Most of these models are leptonic, e.g. internal plus external forward shock 
models were proposed for individual bursts, e.g. \cite{DePasquale+10,Corsi+10}, etc.
Broader formulations have attempted to cover \lat bursts in general, e.g.
\cite{Ghisellini+10}, based on the first three \lat bursts, argued that the
GeV extended emission decays roughly as $F \propto t^{-1.5}$, with \lat spectra of
approximate form $F_E\propto E^{-1}$ which did not evolve strongly. This, they
proposed, can be due to a {\it radiative} (fast cooling) forward shock, where 
the Lorentz factor evolves as $\Gamma \propto t^{-3/7}$ and the luminosity as 
$L\propto T^{-10/7}$. To satisfy the fast cooling ($t_{cool} \leq t_{dyn}$) 
condition in the long-lasting afterglow phase, they proposed that high energy
photons back-scattered towards the source  would provide copious  $e^\pm$ pairs, 
enhancing the cooling. Without going into details, they argued that the external 
shock (GeV) emission would naturally start later than the prompt MeV emission 
(e.g. from internal shocks or similar origin; see also \cite{Meszaros+94gev}).

Alternatively, assuming different power laws through the error bars of the same
burst data, an {\it adiabatic} forward shock may also provide a fit
\cite{Kumar+09fs,Kumar+10fsb}. In this scenario the rough similarity of the spectra 
and decay slopes in several bursts might be due to the GeV emission being above the 
characteristic synchrotron {cooling and peak} energies $E_c,~E_m$, where the 
behavior is insensitive to the initial $\Gamma$ and to the external density $n$, leading 
to a decay $F\propto t^{-1.2}$ typical of adiabatic forward shocks.  These authors 
obtain a range of external densities for which the external shock emission is 
not dominated by the high energy extension of the prompt. Interestingly, these 
include solutions where the pre-shock magnetic field need not be much greater 
than interstellar fields, $B_{ext}\gtrsim 10~\mu$G, to yield after shock compression
a synchrotron flux comparable to that observed, provided the external densities
are uncommonly low, $n_{ext} \lesssim 10^{-5}$ cm$^{-3}$, which may be difficult
(c.f. also \cite{Piran+10,Li10}).  

The inclusion of more detailed  physics can explain some of the departures 
from the simplest forward shock models seen in the data. For instance \cite{Wang+10kn}
show that Klein-Nishina (KN) effects on the SSC upscattering can 
alter the synchrotron source photon flux in a time-dependent manner. The scattering 
Y-parameter is initially low at GeV energies due to KN effects, and weak at X-ray 
energies, leading to strong GeV synchrotron and weak X-ray/optical emission. 
Later the KN effect weakens, Y increases and the GeV synchrotron light curve 
decays more steeply than the $t^{-1.2}$ of  simple adiabatic shocks. Inclusion of
these effects also shows \cite{He+11} that in GRB 090510 the GeV emission in the 
first 4-5 s must have an origin other then the simple forward shock, probably 
being related to the prompt emission.

The bulk Lorentz factors $\Gamma$ obtained from $\gamma\gamma$ pair production
opacity arguments in the prompt emission of bright \lat GRB models such as
the above turned out to be larger than previously expected, $500 \lesssim 
\Gamma \lesssim 10^3$)~\cite{Abdo+09_GRB080916C, Ackermann+10_GRB090510,
Abdo+09_GRB090902B, Ackermann+11_GRB090926A}. These constraints were based
on simple one-zone models, however, and the value depends on model details 
(e.g.~\cite{Baring06,Granot+08}), which are vague at present. Nonetheless, fairly 
general arguments lead to the conclusion \cite{Zou+11} that for the large luminosities 
$L_{\gamma, \rm iso}\gtrsim 10^{54}$ erg~s$^{-1}$ inferred in \lat GRBs, photons 
of energy $E_\gamma \gtrsim 10\GeV$ cannot emerge from radii much smaller than 
$r_{\gamma\gamma}\sim 10^{15}\cm$ (e.g. \cite{Beloborodov10pn}).  The two 
highest-energy (observer frame) photons of 13.2~GeV and 33.4~GeV, detected  from 
from GRB~080916C at $z=4.35$ and GRB~090902B at $z=1.82$, provide through
$\gamma\gamma$ pair production opacity arguments the  best constraints so far~
\cite{Abdo+10_EBL} on models of Extragalactic Background Light (EBL) predicting 
a high optical-UV intensity.

Perhaps the most exciting, or exotic, consequence of the observed delays
between the \lat GeV emission and the \gbm MeV emission is that it can
be used to set robust constraints on effective field theory formulations
of quantum gravity. In particular it rules out a first order dependence on
$(E_\gamma/E_{\rm Planck})$ of any Lorentz invariance violating (LIV) terms,
using GRB~090510 data~\cite{Abdo+09_GRB090510}.

\section{More Detailed Leptonic Models}
\label{sec:gev2}

Attempts at explaining the observed GeV-MeV time delays in the context of simple 
one-zone leptonic Synchrotron-Self-Compton (SSC) mechanisms generally encounter 
difficulties, e.g. \cite{Ackermann+10_GRB090510,Asano+11}. This is the main 
observational motivation for considering two- or multiple zone models, 
typically involving an inner softer source  whose photons are up-scattered by 
electrons in a different region  further out. One example \cite{Toma+09} invokes 
upscattering of X-ray photons from the GRB jet's waste-heat cocoon by internal 
shock electrons. The delayed photons arise from the high latitude brighter 
emission and since the cocoon Lorentz factor is much smaller than that of the
jet, the emission time from the cocoon is delayed relative to that of the jet 
internal shock emission.  This results in two spectral components, but in
some parameter ranges it can also mimic a single Band component.
\begin{figure}[htb]
\begin{minipage}{0.65\textwidth}
\includegraphics[width=0.95\textwidth,height=2.0in,angle=0.0]{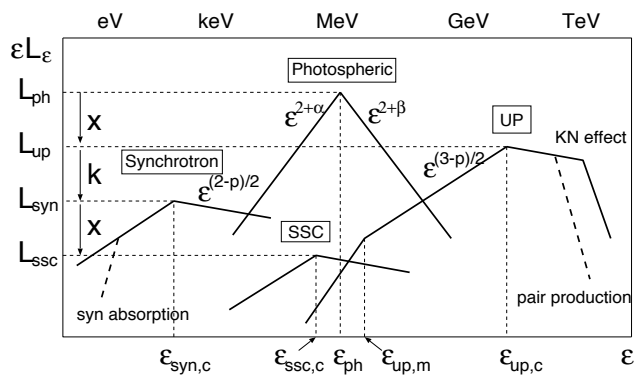}
\end{minipage}
\begin{minipage}[t]{0.3\textwidth}
\vspace*{-1.0in}
\caption{Schematic broadband spectrum for a leptonic photospheric-internal shock 
model of the MeV to GeV emission of GRB (see text) \cite{Toma+11}.}
\end{minipage}
\label{fig:toma-phot}
\end{figure}

Another source of soft photons may be the jet scattering photosphere, which
in some models is thought to be responsible for the Band (MeV) spectrum. 
Detailed numerical calculations of the keV-MeV spectrum of GRB photospheres 
\cite{Lazzati+10phot,Beloborodov10pn} show that a peak can appear at MeV and a 
non-thermal tail extends into the multi-MeV range. Then, upscattering of such 
photospheric photons by internal shock electrons \cite{Toma+11} can lead to an 
upscattered photospheric (UP) component of GeV photons (see Fig. 3, where $k=
U_{syn}/U_{up}$ is the  ratio of synchrotron to upscattered photospheric emission by 
the internal shock, and $x=L_{up}/L_{ph}$ is the ratio of upscattered photospheric 
to photospheric emission).  This results in a broad-band spectrum 
similar to that seen by the \fermi \lat/\gbm (e.g. Fig.  \ref{fig:090926A-spec}).
Time dependent numerical simulations of jets (e.g. \cite{Lazzati+09}) indicate the
possibility of photosphere and internal shock parameters leading to either an 
overlap of the photospheric and the  UP component (appearing as a simple
Band spectrum), or to a separate and bright UP component, corresponding to the
second spectral  component seen in a fraction of \lat GRBs \cite{Toma+11}.

Numerical calculations have been carried out \cite{Asano+11} of the time-dependent 
spectral changes expected in generic dissipation regions (e.g., internal shocks, 
magnetic dissipation, etc.), incorporating all leptonic processes, have shown  
that one-zone models are inadequate to reproduce the observed GeV-MeV delays. 
On the other hand, a multi-zone model injecting different spectral components
at various initial injection times and angles into a geometrically and physically 
separate scattering zone \cite{Asano+11} provides GeV-MeV time delays 
of the right magnitude. These could be, e.g. as above, a photosphere providing a 
Band spectrum plus a second dissipation zone further out which upscatters these 
photons.

\section{Some Issues about Prompt Emission}
\label{sec:basic}

The simple leptonic synchrotron (plus inverse Compton) interpretation of the 
non-thermal aspect of GRB radiation is the most straightforward, and is used 
in preference to others due to its simplicity.  However, a number of effects can 
modify the simple synchrotron spectrum.  One is that the cooling could be rapid, 
i.e.  when the comoving synchrotron cooling time $t'_{sy}=( 9m_e^3 c^5/ 4e^4 B'^2 
\gamma_e) \sim 7\times 10^8/B'^2\gamma_e ~{\rm s}$ is less than the comoving dynamic 
time $t'_{dyn}\sim r/2c\Gamma$, the electrons cool down to $\gamma_c= 6\pi m_e c 
/\sigma_T B'^2 t'_{dyn}$ and the spectrum above $\nu_c\sim \Gamma (3/8\pi)
(eB'/m_e c)\gamma_c^2$ is $F_\nu \propto \nu^{-1/2}$
\cite{sapina98,ghiscel99}. Also, the distribution of observed low energy spectral
indices $\beta_1$ (where $F_\nu\propto \nu^{\beta_1}$ below the spectral peak)
has a mean value $\beta_1\sim 0$, but for a fraction of bursts this slope reaches
positive values $\beta_1>1/3$ which are incompatible with a simple synchrotron
interpretation \cite{preece00}. Possible explanations include synchrotron
self-absorption in the X-ray \cite{grapisa00} or in the optical range
up-scattered to X-rays \cite{panmes00}, low-pitch angle scattering  or jitter
radiation \cite{medvedev00,medvedev06_jitter}, observational selection biases
\cite{Lloyd+01} and/or time-dependent acceleration and radiation \cite{Lloyd+02}
where low-pitch angle diffusion can also explain high energy indices steeper
than predicted by isotropic scattering. Other models invoke a photospheric
component and pair formation \cite{Meszaros+00phot}, see below.

Other recent alternatives to the simple leptonic prompt emission mechanisms have 
continued to be motivated by concerns about the low radiative efficiency and 
fine-tuning needed to achieve the right  peak energy and spectrum in an internal 
shock synchrotron interpretation.  One proposal is based on a relativistic 
turbulence model \cite{Narayan+09turb,Kumar+09turb},
which argues that  relativistic eddies with Lorentz factors $\gamma_r\sim 10$ in 
the comoving frame of the bulk $\Gamma\gtrsim 300$ outflow survive to undergo at 
least $\gamma_r$ changes over a dynamic time, leading both to high variability 
and better efficiency. Various constraints may however pose difficulties 
\cite{Lazar+09}, while numerical simulations \cite{Zhang+09turb} indicate  that 
relativistic turbulence would lead to shocks and thermalization, reducing it to 
non-relativistic. 
A different dissipation model, entitled ICMART  \cite{Zhang+11} involves a
hybrid magnetically dominated outflow leading to semi-relativistic turbulent 
reconnection. Here a moderately magnetized $\sigma=({B'}^2/4\pi\rho' c^2) \lesssim 
100$ MHD outflow undergoes internal shocks as $\sigma \to 1$, leading to turbulence 
and reconnection which accelerates electrons at radii $r\gtrsim 10^{15}$ cm. These 
involve fewer protons than usual baryonic models, hence less conspicuous 
photospheres, and have significant variability, and the efficiency and spectrum  
are argued to have advantages over the usual synchrotron internal shocks 
(see also \cite{Murase+11}.
General issues of the acceleration of high-$\sigma$ relativistic flows were
considered by \cite{Lyutikov10}, while more detailed models were discussed by
\cite{Komissarov+10b,Granot+10,Granot11b}.

\section{Hadronic GRB Models}
\label{sec:hadronic}

\begin{figure}[htb]
{\includegraphics[width=10cm]{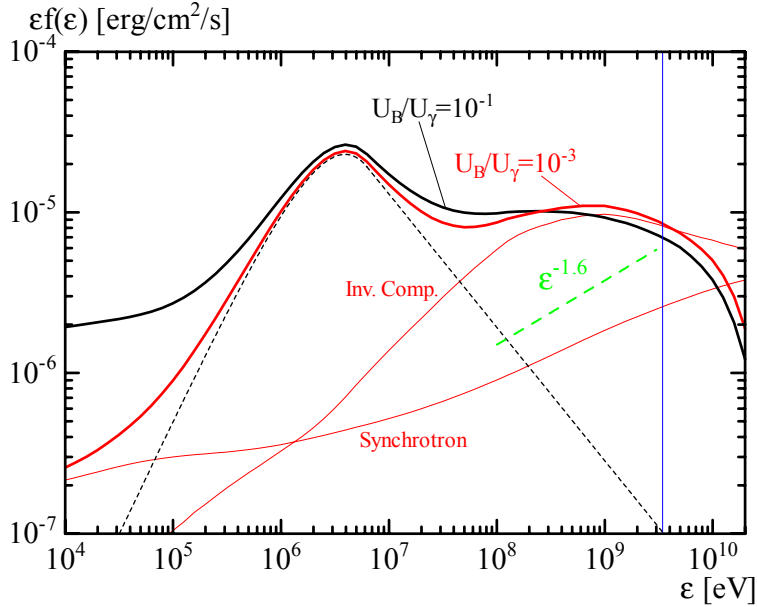}}
\caption{$\nu F_{\nu}$ spectra (bold curve) from a hadronic model of GRB 090510.
The red curve is for $U_B/U_\gamma=10^{-3}$ and $L_{\rm p}/L_\gamma=200$,
and the black curve is for $U_B/U_\gamma=10^{-1}$  and $L_{\rm p}/L_\gamma=30$.
Fine red curves denote separately pair synchrotron and inverse Compton from
hadronic cascade secondaries, without absorption effects.  The fine dashed curve 
is the Band component, the green dashed line is an approximate spectral slope 
of the GeV photons, the vertical blue line denotes the energy of the 3.4 GeV photon 
\cite{Asano+09b}.}
\label{fig:090510-Asano09}
\end{figure}

If GRB jets are baryon loaded, the charged baryons are likely to be co-accelerated
in shocks, reconnection zones, etc., and hadronic processes would lead to both
secondary high energy photons and neutrinos. Monte Carlo codes have been developed  
to model hadronic effects in relativistic flows, including $p,\gamma$ cascades, 
Bethe-Heitler interactions, etc.  E.g., one such code \cite{Asano+09grb,Asano+09-090510} 
was used to calculate the photon spectra in GRBs from secondary leptons resulting from
hadronic interactions following the acceleration of protons in the same shocks
that accelerate primary electrons. The code uses an escape probability formulation
to compute the emerging spectra in a steady state, and provides a detailed
quantification of the signatures of hadronic interactions, which can be compared
to those arising from purely leptonic acceleration. Spectral fits of the \fermi \lat
observations of the short GRB 090510 were modeled by \cite{Asano+09b} as electron 
synchrotron for the MeV component and photohadronic cascade radiation for the 
GeV distinct power law component (Fig. \ref{fig:090510-Asano09}). 

Furthermore, since acceleration as well as cascade development can take additional 
time, even one-zone models might in principle lead to GeV-MeV delays, e.g. in a 
model \cite{Razzaque+10} where the prompt MeV is electron synchrotron and the GeV is 
due to proton synchrotron, whose cooling time brings down the typical photon 
energy into the GeV on the delay timescale, and the electron plus proton synchrotron
emission merge into a single Band functione is an approximate spectral slope
of the GeV photonsn

Hadronic interactions can also have interesting implications for GRB optical prompt
flashes. E.g., as discussed in \cite{Asano+10optex}, besides the usual Band 
MeV spectrum produced by conventional leptonic mechanisms, the acceleration
of hadrons leads to secondaries whose radiation results not only in a high energy 
component but also in a prompt bright optical emission from secondary
synchrotron. This could, in principle,  explain the observed ``naked eye" 5th 
magnitude flash of GRB 080319B, e.g. \cite{Racusin+08-080319}.

Hadronic binary collisions in baryon-loaded jets may also be important, both for 
producing efficient kinetic energy dissipation and for shaping the photon spectrum. 
This is because the baryons will consist (largely) of both protons ($p$) and neutrons 
($n$), especially if heavy elements are photo-dissociated. The protons are coupled to 
the radiation during the acceleration phase but the neutrons are carried along only
thanks to nuclear ($p,n$) elastic collisions, whose characteristic timescale at some 
point becomes longer then the expansion time. At this point the $p$ and $n$ relative
drift velocity $v$ approaches $c$, leading to the collisions becoming inelastic, $p+n 
\to \pi^+,\pi^0$, in turn leading to positrons, gamma-rays and neutrinos 
\cite{Bahcall+00pn}. Such inelastic (p,n) collisions can also arise in jets where 
the bulk Lorentz factor is transversely inhomogeneous \cite{Meszaros+00gevnu}, 
e.g. going from large to small as the angle increases, as expected intuitively 
from a jet experiencing friction against the surrounding stellar envelope. 
In such cases, the neutrons from the slower, outer jet regions can diffuse
into the faster inner regions, leading to inelastic (p,n) and (n,n)  collisions
resulting again in pions. An interesting consequence of either radial or tangential
(n,p) drifts is  that the decoupling generally occurs below the scattering photosphere, 
and the resulting positrons and gamma-rays deposit a significant fraction of the 
relative kinetic energy into the flow, reheating it \cite{Beloborodov10pn}. 
Internal dissipation below the photosphere has been advocated, e.g.  
\cite{Rees+05phot} to explain the MeV peaks as quasi-thermal photospheric 
peaks \cite{Ryde+10,Peer+11}, while having a large radiative efficiency. Such internal 
dissipation is naturally provided by (p,n) decoupling, and numerical simulations 
\cite{Beloborodov10pn} indicate that a Band spectrum and a high
efficiency is indeed obtained, which remains the case even when the flow
is magnetized up to $\vareps_B =2$ \cite{Vurm+11}, while keeping the dynamics
dominated by the baryons. These numerical results were obtained for nominal cases
based on a specific radial (n,p) velocity difference, although the phenomenon
is generic.

\section{Magnetic GRB Models}
\label{sec:mag}

Magnetically dominated (or Poynting dominated) GRB jets fall into two categories, 
one where baryons are absent or dynamically negligible, at least initially, and 
another where the baryon load is significant although dynamically sub-dominant
relative to the magnetic stresses. 

The baryon-free Poynting jet models resemble pulsar wind models, except for 
being jet-shaped, as in AGN baryon-poor models.  The energy requirements of 
GRB (isotropic-equivalent luminosities $L_\gamma \gtrsim 10^{52}$ erg s$^{-1}$)
require magnetic fields at the base in excess of $B\sim 10^{15}$ G, which can
be produced by shear and instabilities in an accreting torus around the BH.
The energy source can be either the accretion energy, or via the magnetic
coupling between the disk and BH, extraction of angular momentum from the 
latter occurring via the Blandford-Znajek mechanism \cite{bz77}. The stresses in
in this type of model are initially magnetic, involving also pairs and photons, 
and just as in purely hydro baryon-loaded models they lead to an initial Lorentz 
factor growth $\Gamma\propto r$ up to a pair annihilation photosphere 
\cite{Meszaros+97poynting}.  This provides a first radiation component, typically 
peaking in the hard X-ray to MeV, with upscattering adding a high energy power law. 
Internal shocks are not expected beyond this photosphere, but an external shock 
provides another IC component, which reaches into the GeV-TeV range.  Such Poynting 
dominated models \cite{Lyutikov+03} may also be characteristic of Pop. III GRBs, 
whose X-ray and GeV spectra may be detectable \cite{Meszaros+10pop3}.
Both the `prompt' emission and the longer-lasting afterglows \cite{Toma+11pop3} 
of such Pop. III GRBs should be detectable with the BAT or XRT on {\it Swift} or 
the GBM on {\it Fermi}. On {\it Swift}, image triggers may be the best way to detect 
them, and some constraints on their rate are provided by radio surveys. 
They are expected to have GeV extensions as well,
but redshift determinations need to rely on L-band or K-band spectroscopy.

The baryon-loaded magnetically dominated jets have a different acceleration 
dynamics than the baryon-poor magnetic jets or the baryon dominated hydrodynamic
jets: whereas both the latter accelerate initially as $\Gamma \propto r$ and 
eventually achieve  a coasting Lorentz factor $\Gamma_f \sim L_\gamma /{\dot M}c^2$, 
the baryon-loaded magnetically dominated jets have a variety of possible 
acceleration behaviors, generally less steep than the above. In the simplest 
treatment of a homogeneous jet with transverse magnetic field which undergoes
reconnection,  the acceleration is $\Gamma\propto r^{1/3}$ \cite{Drenkhahn+02,
Meszaros+11col}, while in inhomogeneous jets where the magnetic field and the
rest mass varies across the jet the average acceleration ranges from 
$\Gamma\propto r^{1/3}$ to various other power laws intermediate between this
and $\Gamma \propto r$ \cite{McKinney+11recon,Metzger+11mgr}. 
Few calculations have been made \cite{Giannios+07spec} of the expected (leptonic) 
spectral signatures in the simpler magnetized outflows, typically in a one-zone 
steady state approximation. 

The photon spectral signatures of a magnetically dominated, baryon loaded
leptonic + hadronic  GRB model involving nuclear collisions has been calculated 
by \cite{Meszaros+11col}. This uses a realistic transverse structure of a fast 
core-slow sheath. The analytical results indicate that the transverse neutron 
collisions become most effective, resulting in GeV photons at radii from which 
the observer-frame time delay relative to the photospheric MeV photons is 
appropriate to explain the observed \fermi time lags.
The purely leptonic (SSC, EIC) time delays and spectral components of such
a baryon-loaded magnetic model, in the absence of drifts and transverse
gradients, have been calculated  by \cite{Bosnjak+11}, leading to results
in the range observed by {\it Fermi}.

\section{High Redshift GRBs}
\label{sec:highz}

GRBs are being identified at increasingly large redshifts, e.g. GRB080913 at $z=6.7$ 
\cite{Greiner+09}, GRB090423 at $z=8.2$ \cite{Tanvir+09,Salvaterra+09} (through
spectroscopy), and GRB 090429B has been ascribed a photometric redshift of $z\simeq 
9.4$ \cite{Cucchiara+11}.
It is possible that even  more distant objects than these have already been detected 
in the gamma- and X-ray detectors of {\it Swift} and {\it Fermi}, although in the absence 
of a specific (optical/IR or other) redshift signature one cannot at present be sure of it,
redshift diagnostics being increasingly harder to obtain in this range.  The above
discoveries do, however, indicate that the prospect of eventually reaching into the 
realm of  Pop. III objects is becoming increasingly realistic.  Pop. III  GRBs may arise 
from very massive, metal-poor stars whose core collapses to a $100-500 \msun$ 
black hole \cite{Komissarov+10}.  The jets are likely to be Poynting-dominated,  
being powered by the Blandford-Znajek mechanism. The expansion dynamics and the 
radiation arising from such very massive Poynting jet GRBs was discussed by 
\cite{Meszaros+10pop3}. At typical redshifts $z\sim 20$ this implies a ``prompt" 
emission extending to $\lesssim$ 1 day which should be detectable by \swift or 
\fermi, being most prominent initially around 50 keV due to the jet pair photosphere, 
followed after a similar time interval by an external shock synchrotron component 
at a few keV and an inverse Compton component at $\gtrsim$ 70 GeV \cite{Toma+11}.

\section{Non-photonic GRB Signals}
\label{sec:nonphot}

The two main non-photonic signals that may be expected from GRBs are gravitational
waves (GW) and high energy neutrinos (HENUs). The most likely GW emitters are short 
GRBs \cite{Centrella+11gwrev}, if these indeed arise from  merging compact objects 
\cite{Gehrels+09araa}. The rates in advanced LIGO \cite{ALIGO} and VIRGO may be at 
least several 
per year \cite{Leonor+09-ligogrb}. Long GRBs, more speculatively, might be detectable 
in GWs if they go through a magnetar phase \cite{Corsi+09mag}, or if the core collapse 
breaks up into substantial blobs \cite{Kobayashi+03gwgrb}; more detailed numerical 
calculations of collapsar (long) GRBs lead to GW prospects which range from 
pessimistic \cite{Ott+11gwcoll} to modest \cite{Kiuchi+11gwcoll}.

High energy neutrinos may be expected from baryon-loaded GRBs if sufficient protons 
are co-accelerated in the shocks. The most widely considered paradigm involves
proton acceleration and $p\gamma$ interactions in internal shocks, resulting in
prompt $\sim 100$ TeV HENUs \cite{Waxman+97grbnu,Murase+06grbnu}. Other interaction 
regions considered are external shocks, with $p\gamma$ interactions on reverse shock UV
photons leading to EeV HENUs \cite{Waxman+00nuag}; and pre-emerging or choked jets 
in collapsars resulting in HENU precursors \cite{Meszaros+01choked}.
An EeV neutrino flux is also expected from external shocks in very massive Pop. III 
magnetically dominated GRBs \cite{Gao+11pop3nu}. Current IceCube observations 
\cite{Ahlers+11-grbprob,Abbasi+11-ic40nugrb} are putting significant constraints on 
the internal shock neutrino emission model, with data from the full array still to be
analyzed.

\section{GRBs and CTA}
\label{sec:grbcta}

Measurements by \lat have led to the conclusion that at least a fraction of GRB
are emitting (in their own rest frame) photons in the energy range of at
least up to $30-90$ GeV. A list of Fermi LAT detections \cite{Omodei+11} 
of maximum observer-frame photon energies and corresponding redshifts 
($E_{\gamma,obs},z$) is $(13.2,4.35),(7.5,3.57),(5.3,0.74),(31.3,0.90),
(33.4,1.82)$, $(19.6,2.10), (2.8,0.897),(4.3,1.37)$.
Two things emerge from this list: even $z>4$ bursts can produce $E_\gamma >10$
GeV photons at the observer, and some $z\sim 1$ bursts can produce $E_\gamma >
30$ GeV photons at the observer. This is highly encouraging for CTA, whether
for a 'baseline' threshold of 25 GeV or for an `optimistic' threshold of 10 GeV
envisaged in recent CTA reviews \cite{Bouvier+11-CTA}. The rate of detection
by CTA is estimated by \cite{Bouvier+11-CTA} to be $0.7-1.6$ per year, based on 
the rate of Swift triggers (while GBM triggers on Fermi are more frequent, their
positional accuracy is poorer). This estimated CTA rate is uncertain, since 
work continues on the evaluation of the actual fraction of bursts which
emit in the GeV range, relative to those which are detected below 100 MeV
\cite{Guetta+11-latgbm,Beniamini+11-latgbm}. As of February 2011, in 2.5 years,
Fermi LAT detected 4 bursts at energies $>10$ GeV (or 20 at $>0.1$ GeV) out of 
some 700 bursts detected by Fermi GBM at $E<100$ MeV. This very small fraction
of the total ($\lesssim 1\%$) of course is in part due to the size constraints 
under which space detectors must operate. The question is what is the real fraction 
of GRBs, whether long or short, which emits significantly at $\gtrsim 10-20$ GeV. 

In the usual internal shock model of prompt emission, the intra-source
$\gamma\gamma$ absorption typically prevents photons in excess of a few GeV
to emerge \cite{Papathanassiou+96is,Pilla+98is}, unless the bulk Lorentz 
factor is above $\sim 700$ \cite{Razzaque+04gev}.
For photospheric models of the prompt emission, e.g. \cite{Beloborodov10pn},
photons in excess of 10 GeV can escape the source from radii $r_{\gamma\gamma}
\sim 10^{15}$ cm, and such radii are also inferred phenomenologically from 
one-zone analyses of the Fermi data on GRB. For the purposes of CTA, however,
it is the afterglow, or extended, GeV emission that is of most interest.
In the standard external shock scenario, the compactness parameter is smaller,
and inverse Compton scattering is expected to lead to multi-GeV and TeV photons 
\cite{Meszaros+94ext,Meszaros+94gev}, the details depending on the electron 
distribution slope and the radiative regime (e.g. slow or fast cooling). This 
scenario is thought to be responsible for the afterglows of GRB \cite{Meszaros+97ag}, 
and is also thought to be responsible for the extended GeV emission observed 
by LAT so far \cite{Ghisellini+10,Kumar+10fsb,Wang+10kn,Zhang+11-latgrb}, etc.  
Of course, propagation in the intergalactic medium from high redshifts leads to 
additional $\gamma\gamma \to e^\pm$ interaction with the extragalactic background 
light, or EBL  \cite{Coppi+97ebl,Finke+10ebl,Primack+11ebl}, the threshold for 
which depends on the photon energy and the source redshift. 

Thus, while TeV emission, if produced, is mainly expected to be detectable from 
$z\lesssim 0.5$, the 10-30 GeV emission should be detectable from higher redshifts, 
as verified by the measurements mentioned above. Therefore, in the GeV range the 
detectability is dictated by the source physics, the source rate and the immediate 
source environment. The source rate, based on MeV observations, is well constrained
\cite{Gehrels+09araa}, while the effects of the near-source environment can be 
reasonably parametrized (e.g. \cite{Gilmore+10gamgam}). The source physics, however, 
offers larger uncertainties. This is because in an external shock model the
simple synchrotron self-Compton (SSC) model can be additionally complicated by 
the scattering of photons arising at other locations, in particular well inside 
the external shock, e.g. from the jet photosphere \cite{Toma+11}, or from an inner 
region energized by continued central engine activity \cite{Wang+06gevflare}.
Similar uncertainties about the soft photon source and location would affect
hadronic cascade models.
Observationally, in some cases the GBM high energy spectral slopes are steep 
enough not to expect much GeV emission from their extrapolation \cite{Omodei+11}, 
while in other cases the LAT spectrum shows a cutoff or turnover, e.g. in
GRB 090926B \cite{Ackermann+11_GRB090926A}. Nonetheless, with all things considered, 
the estimate \cite{Bouvier+11-CTA} of $0.7-1.6$ CTA detections per year appears to 
be a conservative lower limit.

I am grateful to N. Omodei, S. Razzaque and P. Veres for valuable discussions.
This research was supported in part by NASA NNX08AL40G and NSF PHY-0757155.

\end{document}